\documentclass[]{interact}

\usepackage{epstopdf}
\usepackage{listings}
\usepackage[caption=false]{subfig}
\lstdefinelanguage{Smalltalk}{
  morekeywords={self,super,true,false,nil,thisContext},
  sensitive=true,
  morecomment=[s]{"}{"},
  morestring=[d]',
}
\usepackage{caption}
\usepackage[all]{nowidow}
\usepackage{mdframed}
\usepackage{xparse}
\usepackage{minted}
\usepackage{flushend}
\usepackage{hyperref}
\usepackage{color}
\usepackage{booktabs}
\usepackage{xspace}
\usepackage{multirow}
\usepackage{colortbl}
\usepackage{ragged2e}
\usepackage{courier}
\usepackage{textcomp} 
\usepackage{listings}
\usepackage{varwidth}
\usepackage[normalem]{ulem}
\usepackage{paralist}
\usepackage{booktabs}
\usepackage{mdframed}
\usepackage{adjustbox}
\usepackage{caption}
\usepackage{subcaption}
\usepackage{makecell}
\usepackage{orcidlink}

\newboolean{showedits}
\setboolean{showedits}{false} 
\ifthenelse{\boolean{showedits}}
{
	\DeclareRobustCommand{\ins}[1]{{\color{blue}#1}} 
	\DeclareRobustCommand{\del}[1]{{\color{red}\sout{#1}}} 
	\newcommand{\chg}[2]{\textcolor{red}{\sout{#1}}{\ra}\textcolor{blue}{\uline{#2}}} 
}{
	\newcommand{\ins}[1]{#1} 
	\newcommand{\del}[1]{} 
	\newcommand{\chg}[2]{#2}
}
\newboolean{showcomments}
\setboolean{showcomments}{true}

\ifthenelse{\boolean{showcomments}}{%

  \DeclareRobustCommand{\nbc}[3]{%
    {\colorbox{#3}{\bfseries\sffamily\scriptsize\textcolor{white}{#1}}}%
    {\textcolor{#3}{\sf\small$\blacktriangleright$\textit{#2}$\blacktriangleleft$}}%
  }%

}{%

  \DeclareRobustCommand{\nbc}[3]{}%
  \DeclareRobustCommand{\ins}[1]{#1}
  \DeclareRobustCommand{\del}[1]{}
  \DeclareRobustCommand{\chg}[2]{#2}

}

\newcommand{\eg}{{e.\,g.},\xspace} 
\newcommand{\etal}{et al.\xspace}
\newboolean{isblinded}
\setboolean{isblinded}{false}
\ifthenelse{\boolean{isblinded}}
{\newcommand\blind[1]{BLINDED\xspace}}
{\newcommand\blind[1]{#1\xspace}}

\usepackage[english]{babel}
\usepackage{listings}
\usepackage{xparse}
\definecolor{light-gray}{gray}{0.97}

\newcommand{\myparagraph}[1]{\vspace{0.1cm}\noindent \textit{\textbf{#1.}}}

\usepackage[all]{nowidow}
\usepackage[numbers,sort&compress,merge]{natbib}
\bibpunct[, ]{(}{)}{,}{n}{,}{,}

\theoremstyle{plain}

\theoremstyle{definition}

\theoremstyle{remark}

\begin{document}


\title{In-Situ Immersive Analytics Authoring through Ergonomic Keyboard Support}

\author{
\name{Leonel Merino\textsuperscript{a}\textsuperscript{b}\orcidlink{0000-0002-5396-487X}\thanks{CONTACT Leonel Merino. Email: leonel.merino@uc.cl}, Bego\~{n}a Juli\'{a}-Nehme\textsuperscript{a}\textsuperscript{b}\orcidlink{0000-0003-4888-9868} and Santiago Viana\textsuperscript{a}\orcidlink{0009-0000-9343-9095}}
\affil{\textsuperscript{a}School of Design, Faculty of Architecture, Design, and Urban Studies, Pontificia Universidad Católica de Chile, Santiago, Chile; \textsuperscript{b}Engineering Design (DILAB), School of Engineering, Faculty of Engineering, Pontificia Universidad Católica de Chile, Santiago, Chile}
}

\maketitle

\begin{abstract}
Immersive analytics uses augmented reality (AR) to integrate data analysis and authoring within physical environments\del{ of users}. However, \ins{extensive} text entry \ins{required for immersive analytics authoring} remains a \ins{fundamental} challenge \ins{in AR}, as \ins{popular} natural user interfaces often hinder \chg{efficient}{expressive} input. This paper presents the Body-Supported Keyboard (BSK), an ergonomic system that allows the mobile use of a Bluetooth keyboard in AR. We conducted a controlled study with 20 participants to compare the BSK with a standing desk during text transcription and a mobile AR scenario. The results showed slightly higher error rates but comparable task completion times. Participants reported comfort improvements during mobile use and positive usability ratings (mean SUS = 74.5). The BSK allows users to move freely and maintain stable postures while authoring in AR. In general, the findings show evidence of the potential for body-supported input to enhance expressive and ergonomic workflows in immersive analytics and emphasize the importance of comfort and mobility in the design of AR authoring tools.
\end{abstract}


\section{Introduction}
\label{sec:introduction}
\ins{Immersive Analytics (IA) explores how immersive technologies such as Augmented Reality (AR) and Virtual Reality (VR) can enhance data understanding and decision-making by embedding visualizations directly within the environments where users work and reason~\cite{elsayed2016situated}. 
In doing so, IA aims to bridge abstract data analysis with the physical contexts in which data are produced and acted upon~\cite{Card97a,Marr18a}.
Achieving this goal places strong demands on the interaction techniques used for authoring, manipulation, and sustained engagement within immersive environments.

To support this vision, a variety of authoring toolkits~\cite{Meri20b,Sica19a,Cord19a,Cord17a,Dona14a,nebeling2020mrat,yim2018niwviw,Saif18a,butcher2019vria} have been proposed to enable the creation of immersive analytics experiences without requiring developers to build systems from scratch. Despite this progress, most current workflows for immersive analytics authoring remain strongly desktop-centered, posing a barrier to iterative and in-situ design. 
Given extensive text-entry requirements, visualizations are typically designed and programmed on conventional workstations and only later deployed to immersive environments such as AR headsets~\cite{butcher2019vria}. This separation introduces friction, interrupting iterative authoring, debugging, and sensemaking, and undermining the benefits of immersion~\cite{kazemi2024human}.

This limitation becomes particularly acute for professional authoring tasks, 
such as programming visualization logic, defining data transformations, or configuring interaction behaviors, where expressiveness is essential. 
While graphical and natural user interfaces can support low-threshold interaction, they are insufficient for such tasks, making expressive and sustained text entry a critical bottleneck for in-situ immersive analytics authoring. 

Existing AR and VR text-entry techniques predominantly rely on mid-air interaction, surface-projected keyboards, or unfamiliar input devices. While these approaches enable basic text input, they often incur substantial penalties in typing speed, accuracy, fatigue, or ergonomic comfort, particularly during prolonged use. As a result, traditional physical keyboards remain difficult to replace in mobile or spatially dynamic AR scenarios~\cite{cuaresma2013study}.

In this paper, we address this gap by exploring whether body-supported physical input can enable expressive, mobile authoring workflows in augmented reality without sacrificing typing performance or ergonomic comfort. We introduce the Body-Supported Keyboard (BSK), an ergonomic system that secures a standard Bluetooth keyboard to the user’s body, preserving tactile familiarity while allowing users to remain mobile and engaged with their physical surroundings. By supporting a conventional input device through an ergonomic, body-supported configuration, the BSK introduces a novel approach to text entry in AR that facilitates expressive in-situ authoring while lowering barriers to user understanding and adoption.

We evaluate the BSK through a controlled user study that compares typing performance, postural discomfort, and perceived usability against a conventional standing desk setup. Our results show that the BSK preserves typing performance while supporting dynamic posture changes and reducing physical strain during in-situ authoring tasks. These findings show that body-supported input can offer a viable alternative to mid-air or surface-based text entry, particularly for immersive analytics scenarios that demand prolonged and expressive text input.

Unlike prior approaches that rely on mid-air input, surface-projected keyboards, or fixed physical desks, the BSK introduces a body-supported configuration that preserves tactile familiarity while enabling mobile, in-situ authoring. The novelty lies not in the keyboard itself, but in its ergonomic integration with the body to support expressive programming workflows under movement, an aspect not addressed by existing AR text-entry systems.}

\myparagraph{Contributions} 
This paper investigates the feasibility and ergonomic implications of supporting programming workflows in augmented reality using a familiar text input device. The main contributions are as follows.

\begin{itemize}
    \item The design and implementation of the BSK system that enables the entry and programming of mobile text within AR environments.
    \item A controlled user study that compared typing performance and postural discomfort when using the BSK versus a conventional standing desk.
    \item Empirical evidence demonstrating that the BSK supports dynamic posture changes and natural movement during in-situ authoring tasks, along with overall usability perception of the system.
    \item A discussion of the role of ergonomic design in immersive analytics toolkits to improve comfort, usability, and technology acceptance.
\end{itemize}

We believe that this work will inform both researchers and practitioners. Designers of immersive analytics systems can integrate ergonomics more systematically into their tools, while researchers can build on this evidence to explore hybrid input methods and longer-term field evaluations.

\section{Background}
\ins{Prior work relevant to this paper spans immersive analytics authoring, interaction techniques for text entry in AR and VR, ergonomic considerations in immersive systems, and factors influencing technology acceptance. Together, these strands of research highlight both the opportunities and the practical challenges of supporting expressive, in-situ authoring workflows in immersive environments.

\subsection{Immersive Analytics Authoring in Situated Contexts}
We organize existing research on immersive analytics authoring around two complementary perspectives. We first summarize prior work on situated visualizations and authoring tools that support in-situ immersive analytics. We then examine the interaction modalities used in these systems, with particular attention to text entry constraints that affect expressive authoring in AR and VR.

\subsubsection{Situated Visualizations and Authoring Toolkits}
Immersive Analytics combines data visualization with immersive technologies to support situated and embodied data analysis~\cite{elsayed2016situated}. Within this space, situated visualizations play a key role by embedding visual representations directly into the user’s physical environment, often in direct relation to real-world objects, locations, or processes~\cite{martins2022augmented}. AR-based situated visualizations have been shown to support contextually grounded decision-making in domains such as maintenance, logistics, healthcare, and field engineering~\cite{sauter2014decision}.

To enable the creation of such situated immersive analytics experiences, numerous authoring toolkits have been proposed~\cite{Meri20b,Sica19a,Cord19a,Cord17a,Dona14a,nebeling2020mrat,yim2018niwviw,Saif18a,butcher2019vria}. These systems aim to lower development barriers by providing abstractions for data binding, visual mappings, and interaction techniques. However, most still rely on workflows that begin on desktop platforms (\eg Unity) before deployment to immersive devices, limiting support for in-situ experimentation and iterative authoring~\cite{kazemi2024human}.

This workflow gap is closely tied to how immersive authoring toolkits support interaction and input, particularly for tasks that require expressive specification and modification of visualization logic.

\subsubsection{Interaction Modalities and Text Entry Constraints}
Within these authoring toolkits, immersive systems commonly support three interaction modalities: natural user interfaces (NUIs), graphical user interfaces (GUIs), and text-based programming interfaces~\cite{cuaresma2013study,yu2017tap,kazemi2024human}. 
NUIs such as hand gestures, gaze, and speech are often intuitive and easy to learn, while GUIs support rapid configuration through visual controls. 
However, both modalities typically lack the expressive power required for complex data transformations or highly customized visualizations~\cite{mcgill2015dose}.

As a result, text-based programming interfaces remain essential for professional immersive analytics workflows, where authoring tasks frequently involve defining visualization logic, data transformations, and interaction behaviors. Text-based input provide greater expressiveness and extensibility, making them essential for professional workflows. 
Yet, supporting text entry in immersive environments remains challenging. 
A large body of work has explored mid-air keyboards, gaze-based input, raycasting techniques, chorded keyboards, and wearable or handheld devices~\cite{mcgill2015dose,grubert2018text,speicher2018selection,knierim2018physical}. 
While these approaches enable text entry without leaving the immersive environment, they generally suffer from reduced typing speed, higher error rates, increased fatigue, or limited comfort compared to conventional keyboards.

Recent studies illustrate this trade-off. Mid-air QWERTY keyboards typically achieve around 20–30 WPM but are affected by arm fatigue and the absence of passive haptics~\cite{dudley2023probabilistic}. Body- or surface-supported approaches improve stability and accuracy; for example, projecting a keyboard onto the palm achieved 20.18 WPM with low error rates~\cite{darbar2024onarm}. To contextualize these trade-offs across techniques, Table~\ref{tab:typing-comparison} summarizes representative results across text-entry techniques in immersive environments.

\setlength{\tabcolsep}{3pt}
\begin{table*}[ht]
\centering
\tiny
\setlength\tabcolsep{0.5pt}
\begin{tabular}{p{4.1cm} p{4cm} c c p{4.2cm}}
\toprule
\textbf{Study} & \textbf{Text Entry Method} & \textbf{Mean WPM} & \textbf{Error Rate (\%)} & \textbf{Notes} \\
\midrule
Darbar et al. (2024)~\cite{darbar2024onarm} & On-arm QWERTY (AR) & 20.3 & 0.7 & Wearable on-arm keyboard\\
Wang et al. (2023)~\cite{wang2023eyeshaped} & Eye-shaped (VR) & 19.9 & 1.9 & Raycast typing with curved layout\\
Wan et al. (2024)~\cite{wan2024handsfree} & Hands-free (VR) & 8.5--7.8 & $<$3.0 & Dwell and blink-based gaze selection\\
Dudley et al. (2023)~\cite{dudley2023probabilistic} & Mid-air QWERTY (VR) & 25.6--21.5 & \(\sim\)2.25 & Hand-tracked typing in mid-air\\
Fallah \& MacKenzie (2023)~\cite{fallah2023h4vr} & H4VR 4-key (VR) & 4.6--3.6 & $<$2.0 & Ambiguous keyboard using gestures\\
Grubert et al. (2018)~\cite{grubert2018text} & Freehand keyboard (VR) & \(\sim\)15.0 & \(\sim\)4.0--6.0 & Typing in mid-air with hand tracking \\
Knierim et al. (2018)~\cite{knierim2018physical} & Physical keyboard (VR) & \(\sim\)12.0--16.0 & \(\sim\)5.0--8.0 & Different hand representations in VR \\
Speicher et al. (2018)~\cite{speicher2018selection} & Selection-based text entry (VR) & \(\sim\)5.0--10.0 & \(\sim\)5.0--7.0 & Menu and dwell-based selection \\
McGill et al. (2015)~\cite{mcgill2015dose} & Virtual keyboard (VR) & \(\sim\)8.6 & \(\sim\)5.5 & Substantial slowdown \\
McGill et al. (2015)~\cite{mcgill2015dose} & Chorded keyboard (VR) & \(\sim\)6.0 & \(\sim\)5.0--8.0 & Compact input method \\
Vertanen \& Kristensson (2011)~\cite{vertanen2009parakeet} & Dasher selection-based input & \(\sim\)12.0 & \(\sim\)3.0--5.0 & Gaze/gesture continuous input \\
Majaranta \& Räihä (2002)~\cite{majaranta2002twenty} & Eye-gaze typing & \(\sim\)5.0--8.0 & \(\sim\)6.0--10.0 & Dwell-based gaze selection \\
\bottomrule
\end{tabular}
\caption{Comparison of text entry performance across studies evaluating different input methods in immersive environments}
\label{tab:typing-comparison}
\end{table*}

Overall, these findings indicate that tangible or body-supported input can better preserve text-entry performance than purely mid-air techniques, particularly for tasks requiring sustained and precise input~\cite{darbar2024onarm,dudley2023probabilistic}.

\subsection{Ergonomics in Immersive Systems}
Ergonomics is a critical consideration for immersive systems, especially in professional contexts involving prolonged use. In AR and VR, mid-air interaction and head-mounted displays can impose substantial musculoskeletal load, leading to discomfort and fatigue~\cite{kia2023effects,kim2020evaluation,lim2024physical,penumudi2020effects,du2023comfort,ito2021effects}. Physiological studies show increased neck and shoulder strain during AR interactions and highlight the influence of headset mass and center-of-mass on cervical torque~\cite{kia2023effects,kim2020evaluation,astrologo2024}.

These findings motivate the exploration of body-supported input techniques that can reduce static load, improve postural stability, and support longer interaction sessions. Pairing subjective measures with physiological data such as EMG and motion capture has been proposed as a way to better understand ergonomic trade-offs in immersive interaction design.

\subsection{Technology Acceptance}
Beyond performance and ergonomics, adoption of immersive analytics tools depends on user acceptance. The Technology Acceptance Model (TAM)~\cite{davis1989perceived} and the Unified Theory of Acceptance and Use of Technology (UTAUT)~\cite{venkatesh2003user} emphasize perceived usefulness, ease of use, social influence, and facilitating conditions as key factors of sustained adoption. Recent industrial studies further show that ergonomic comfort and familiarity with existing tools strongly influence sustained use of AR head-mounted displays in professional environments~\cite{okeeffe2024ergonomics}.

Despite substantial progress in immersive text-entry techniques, many AR systems still rely on mid-air or surface-projected input. While effective for short interactions, these approaches often reduce typing accuracy, postural stability, and comfort during extended use, negatively impacting perceived usability and acceptance. Prior work consistently reports increased fatigue and reduced confidence when text entry departs from familiar physical input, particularly in mobile contexts.

Taken together, prior work suggests that successful immersive analytics authoring toolkits must balance expressiveness, performance, ergonomic comfort, and familiarity. Our work builds on these insights by examining a body-supported physical keyboard as a means to support expressive, mobile authoring in AR while preserving the tactile and behavioral characteristics of conventional typing.}

\section{Design Study: The BSK system}
To enable programming of immersive analytics visualizations directly within augmented reality environments, it is essential to integrate a text input device that preserves efficiency, expressiveness, and ergonomics. 
\ins{However, existing data input devices for AR environments often reduce typing accuracy and task efficiency, and their adoption is hindered by the learning effort required to master unfamiliar interaction techniques.}
\chg{Although}{In contrast, traditional} Bluetooth keyboards \chg{provide a familiar and}{offer high} efficien\chg{t means of entering text,}{cy and strong familiarity, enabling programmers to transfer existing typing skills into AR environments; however,} carrying and stabilizing them during in-situ or mobile tasks introduces significant design challenges. 
This section describes the development of the \ins{Body-Supported Keyboard (}BSK\ins{)}\del{ system}, a wearable \chg{system designed to allow users to work comfortably and safely while maintaining freedom of movement in the field}{ergonomic support that brings a traditional Bluetooth keyboard into the field, allowing users to work comfortably and safely while remaining mobile.}. 
We first review ergonomic principles and recommendations relevant to computer-based typing tasks and then present the design process and considerations that guided the creation of the BSK prototype.

\ins{\subsection{Design Goals and Requirements}
The BSK was designed around a set of goals and functional and ergonomic requirements aimed at supporting expressive, mobile programming in AR environments. To address the challenges of using a Bluetooth keyboard while in motion, the design process focused on the following key objectives:

\begin{itemize}
    \item Providing a stable platform for secure carrying of the keyboard during use.
    \item Providing a variety of body sizes and shapes through adjustable components.
    \item Supporting the forearms to promote healthy typing posture according to ISO 9241-5 recommendations, specifically maintaining approximately 90$^\circ$ of elbow flexion and keeping the wrists aligned with the forearms while typing.
    \item Allowing freedom of movement, postural adjustments, and performance of common field activities such as walking or climbing stairs.
\end{itemize}}

\subsection{Ergonomic Rationale for Body-Supported Typing}
To inform these requirements, we draw on established ergonomic research on typing and posture. Extensive research has examined the factors that influence health and performance during computer-based work. Typing, in particular, is a repetitive activity that places pressure on the upper extremities and can lead to musculoskeletal disorders. For example, Shiri and Falah-Hassani~\cite{shiri2015computer} identified an association between carpal tunnel syndrome and computer use among office workers. Other studies have highlighted that extreme wrist postures increase the risk of hand and wrist musculoskeletal symptoms~\cite{liu2003relationship,malchaire1996prevalence,viikari1999role}. When combined with repetitive tasks, these postures further increase the likelihood of injury~\cite{bp1997musculoskeletal}.
To mitigate these risks, the ISO 9241-5 standard for ergonomic requirements in office work with visual display terminals (ISO, 2024) recommends adopting a posture in which the elbows are flexed at approximately 90 degrees or slightly more open and the forearms are positioned parallel to the floor while typing. The wrists should remain straight, avoiding excessive extension. To support this posture, users can rest their forearms on a desk surface or rely on adjustable armrests. Previous studies have shown that armrests effectively reduce the physical load on the neck, shoulders, and upper extremities~\cite{feng1997effects,paul1996impact}, and that users often prefer workstations equipped with these supports~\cite{aaras1997postural}.

In addition to seated postures, evidence suggests that typing performance is not adversely affected by standing. For example, Kar and Hedge~\cite{kar2016effects} found that short-term typing tasks performed while standing resulted in higher typing speeds and fewer errors compared to sitting. Although participants reported greater discomfort in the lower body while standing, this discomfort could be alleviated by incorporating movement. Similarly, Fedorowich and Côté~\cite{fedorowich2018effects} observed that standing was associated with increased typing speed and reduced upper body discomfort. Commissaris et al.~\cite{commissaris2014effects} also confirmed that standing does not compromise typing efficiency.

It is also important to consider that maintaining an ergonomic posture can enhance the perception of comfort. Vink and De Looze~\cite{vink2008comfort} emphasized the importance of comfort in product design, noting that the presence of discomfort can contribute to musculoskeletal injuries. They identified three key factors to consider: the duration of exposure, the distinction between short- and long-term (dis)comfort; the subjective perception of comfort; and the identification of design elements that influence comfort.

In general, these findings provide important considerations for designing an ergonomic, portable system to support keyboard use in AR environments. Ensuring neutral wrist and elbow postures, offering adequate forearm support, and allowing opportunities for movement are critical to promoting comfort and reducing the risk of musculoskeletal strain during immersive analytics tasks.

\subsection{Design Process \ins{and Final Prototype}}
\del{To address the challenges of using a Bluetooth keyboard while in motion, we developed the BSK prototype guided by both functional requirements and ergonomic principles. The design process focused on satisfying four key objectives:}
\del{
\begin{itemize}
    \item Providing a stable platform for secure carrying of the keyboard during use.
    \item Providing a variety of body sizes and shapes through adjustable components.
    \item Supporting the forearms to promote healthy typing posture according to ISO 9241-5 recommendations, specifically maintaining approximately 90$^\circ$ of elbow flexion and keeping the wrists aligned with the forearms while typing.
    \item Allowing freedom of movement, postural adjustments, and performance of common field activities such as walking or climbing stairs.
\end{itemize}
}
We carried out an iterative design process involving five students from the Design and Engineering disciplines. 
The development progressed from early sketches to successive physical prototypes, which were tested in team members with various body types to evaluate fit, adjustability, and comfort. 
The design of the armrest adjustment mechanism proved particularly challenging, as it required balancing simplicity with sufficient adaptability while avoiding the use of overly complex components. 
In addition, the keyboard tray was iterated several times to ensure compatibility with different keyboard sizes and to provide adequate stability for typing tasks.

\begin{figure}[t]
    \centering
    \includegraphics[width=\textwidth]{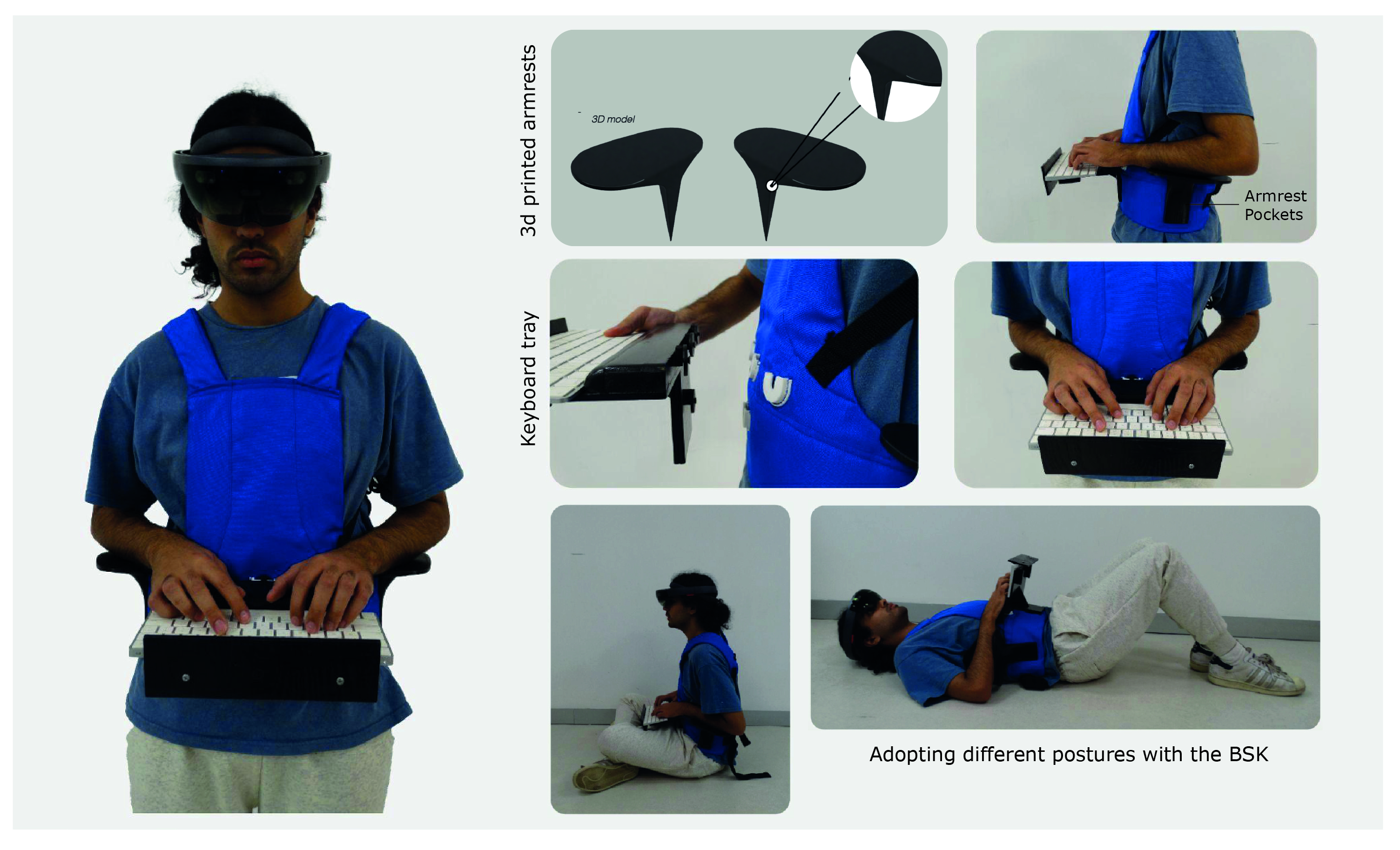}
    \caption{Design and prototype of BSK.}
    \label{fig:bsk}
\end{figure}

The final BSK prototype consisted of several components (Figure~\ref{fig:bsk}) designed to integrate ergonomic support and adaptability for immersive analytics authoring. 
The illustration shows an adjustable fabric vest equipped with plastic buckles and hook and loop (Velcro) fasteners to secure it around the user’s torso, shoulders, and waist. 
From the side, the prototype demonstrates how the keyboard tray attaches to the front. 
The belt component includes armrest pockets on each side that allow users to select the most comfortable position. 
It also features frontal attachment points to connect the keyboard support tray. 
The view shows the vest worn by a user and illustrates how it wraps securely around the torso. 
This configuration helps maintain a neutral posture and allows freedom of movement, including reaching and walking. 

A pair of 3D printed armrests designed with integrated square bases that fit securely into belt pockets. 
These armrests provide stable support for the forearms, reducing strain on the shoulders and wrists during typing tasks. 
The system also includes a 3D printed keyboard tray that can be adjusted to accommodate keyboards of various sizes. 
The tray securely holds the keyboard in place while allowing easy access for typing.

This functional prototype was developed to allow users to maintain ergonomic posture and freedom of movement while typing with a Bluetooth keyboard in mobile or field environments. 
It enabled us to test the feasibility of programming immersive analytics visualizations directly in-situ, without the need to return to a stationary workstation.
\ins{This prototype served as the basis for the controlled and mobile evaluations described in the following user study.}

\section{User Study}
\label{sec:methodology}

To evaluate the performance, comfort, and feasibility of using the BSK for programming immersive analytics visualizations in augmented reality, we conducted a mixed-methods user study combining elements of a controlled experiment and a simulated field assessment. The study was designed to capture both quantitative performance metrics and qualitative observations of real-world use. Specifically, it included two tasks: (a) a controlled comparative experiment in which participants transcribed code using either a standing desk (Task A1) or the BSK (Task A2), enabling direct within-subject comparison of typing performance and physical comfort; and (b) a simulated field scenario (Task B) in which participants moved freely around the laboratory while using the BSK to complete code scripts and manipulate data visualizations. This combination allowed us to assess not only the impact of the BSK on accuracy and efficiency but also its practical usability, comfort, and support for mobility in more realistic conditions. 

This study received ethical approval from the Ethics Committee of the Pontificia Universidad Católica de Chile (CEC CSAH N° 220321014). All procedures were carried out according to the ethical standards of the committee and the principles outlined in the Declaration of Helsinki. Written informed consent for participation in the study was obtained from all participants prior to data collection.

\subsection{Tasks}
The study was structured \chg{around}{following a} two\chg{ tasks designed to assess both stationary and mobile use of the BSK system.}{-stage evaluation strategy designed to (i) establish baseline text-entry performance under controlled conditions and (ii) extend and contextualize these findings in an embodied, mobile AR authoring workflow. Task A quantifies performance and error characteristics of the BSK relative to a conventional standing desk while minimizing confounds such as navigation or search. Task B then uses a mobile, in-situ authoring scenario to examine how the BSK supports movement, posture changes, and repeated short coding actions that are characteristic of situated visualization work.}

\noindent\textbf{Task A: \chg{Comparative}{Baseline} typing performance \ins{under controlled conditions}.}
\ins{Task A served as the performance baseline. }Participants were asked to transcribe a script into the AR system under two conditions: Task A1, using a standing desk with adjustable height. The A1 script is presented in~\autoref{lst:task-a1}; and Task A2, using the BSK. Script A2 is presented in~\autoref{lst:task-a2}). The order of these tasks was counterbalanced among the participants to mitigate possible order effects~\cite{wohlin2012experimentation}. Each script consisted of 833 characters of Smalltalk code and was printed on paper. The scripts were designed to be equivalent in length and difficulty to ensure fair comparison. Participants were instructed to enter the code accurately and were not allowed to make corrections after submission.

\ins{\noindent\emph{Rationale for the Task A scripts.}
The scripts in Listings~\ref{lst:task-a1} and~\ref{lst:task-a2} were selected to approximate realistic visualization authoring actions while remaining suitable for controlled transcription. Both scripts require typical programming operations involved in building visualizations: selecting and filtering data (\eg \texttt{select:} predicates on \texttt{numberOfLinesOfCode}), configuring visual encodings (\eg \texttt{dotShape}, \texttt{color}, \texttt{alpha}, and \texttt{size}), defining mappings to axes (\texttt{x:} and \texttt{y:}), and executing the visualization (\texttt{build}, \texttt{view run}). They also include a small amount of structural complexity (multiple assignments, cascades, blocks, and message sends) representative of actual authoring, without requiring domain knowledge beyond transcription. Importantly, the scripts are comparable in length and overall structure but not identical, reducing memorization effects between A1 and A2 while preserving the same class of authoring operations.}

\begin{lstlisting}[language=Smalltalk, caption={Script Task A1}, label={lst:task-a1}]
| methodsRoassal methodsTrachel greatestAge b ds | 
methodsRoassal := RTObject withAllSubclasses flatCollect: #rtmethods.
methodsTrachel := TRObject withAllSubclasses flatCollect: #rtmethods.
methodsRoassal := methodsRoassal
    select: [ :m | m numberOfLinesOfCode > 50 ].
methodsTrachel := methodsTrachel
    select: [ :m | m numberOfLinesOfCode > 50 ].
b := RTGrapher new.
b extent: 300 @ 300.
ds := RTData new.
ds interaction popup.
ds dotShape circle
    color: (Color blue alpha: 0.3);
    size: 5.
ds points: methodsRoassal.
ds x: [ :m | 100 atRandom ].
ds y: #numberOfLinesOfCode.
b add: ds.
ds := RTData new.
ds dotShape circle
    color: (Color red alpha: 0.3);
    size: 5.
ds points: methodsTrachel.
ds x: [ :m | 100 atRandom ].
ds y: #numberOfLinesOfCode.
b add: ds.
b axisX numberOfTicks: 4.
b axisY numberOfTicks: 4.
b build.
b view run.
\end{lstlisting}

\begin{lstlisting}[language=Smalltalk, caption={Script Task A2}, label={lst:task-a2}]
| methodsRoassal methodsTrachel greatestAge b ds |
methodsRoassal := RTObject withAllSubclasses flatCollect: #rtmethods.
methodsTrachel := TRObject withAllSubclasses flatCollect: #rtmethods.

methodsRoassal := methodsRoassal select: [ :m | m numberOfLinesOfCode < 300 ].
methodsTrachel := methodsTrachel select: [ :m | m numberOfLinesOfCode < 300 ].

greatestAge := (methodsTrachel , methodsRoassal collect: [ :m | m date asUnixTime ]) min.

b := RTGrapher new.
b extent: 300 @ 300.

ds := RTData new.
ds interaction popup.
ds dotShape circle color: (Color blue alpha: 0.3); size: 5.
ds points: methodsRoassal.
ds x: [ :m | ((m date asUnixTime - greatestAge) / 3600) asFloat ] .
ds y: #numberOfLinesOfCode.
b add: ds.

ds := RTData new.
ds dotShape circle color: (Color red alpha: 0.3); size: 5.
ds points: methodsTrachel.
ds x: [ :m | ((m date asUnixTime - greatestAge) / 3600) asFloat ] .
ds y: #numberOfLinesOfCode.
b add: ds.

b build.
b view run.
\end{lstlisting}

\noindent\textbf{Task B: BSK Use Behavior and Comfort.}
\ins{Task B built on the baseline from Task A by placing coding within a mobile AR workflow.} Participants were asked to create three different visualizations. For each visualization, they opened an incomplete predefined script stored in the AR interface~. The script is presented in~\autoref{lst:task-b}. The participants completed it by finding an additional piece of code placed somewhere in the laboratory. To accomplish this, participants had to move around the space to find the correct label containing the missing code segment, transcribe it using the keyboard, execute the completed script, and then position the resulting visualization in a designated area within the environment. \ins{This task explicitly introduces mobility, attention switching between the environment and the AR interface, and repeated short transcription and execution cycles (factors that are absent in Task A but central to in-situ authoring).}

\ins{\noindent\emph{Rationale for the Task B script.}
Listing~\ref{lst:task-b} was designed to reflect a common ``edit--run--inspect'' loop in visualization authoring while keeping each iteration short enough to emphasize embodied interaction rather than sustained transcription. The script includes typical visualization construction steps (dataset assignment, grapher creation, visual encoding via \texttt{barShape}, conditional styling, and execution via \texttt{build} and \texttt{open}). By embedding only a small missing segment per trial, Task B focuses on how the BSK supports practical in-situ use: brief but repeated text entry interleaved with movement and spatial placement. Using three visualizations provides multiple cycles to observe adaptation, posture variation, and comfort trends across repeated mobile authoring actions.}

\begin{lstlisting}[language=Smalltalk, caption={Script Task B}, label={lst:task-b}]
data := #( 10 20 30).
b := RTGrapher new.
b extent: 300 @ 200.
ds := RTData new.
ds interaction popup.
ds points: data.
ds barShape
	width: 8;
	color: (Color green alpha: 0.3);
	if: [ :value | value < 0 ] fillColor: (Color red alpha: 0.3).
b add: ds.
b build.
b open.
\end{lstlisting}

\subsection{Participants}

We recruited a sample of 20 participants to enable within-subject comparative analysis in Task A and observe behavioral and comfort-related aspects of BSK use in Task B. Inclusion criteria required participants to be at least 18 years old and to have a minimum of one year of programming experience. Participants were recruited through social media outreach and direct invitations from the research team. All participants provided their informed consent prior to participating in the study.

\ins{The participant pool primarily consisted of young adults with a technical background, which is consistent with the exploratory and controlled nature of this study. This sample is well suited for an initial evaluation of typing performance, ergonomic effects, and usability under comparable conditions. However, the demographic composition may limit the generalizability of the findings to broader professional populations, such as older users or experienced field technicians.}

\subsection{Apparatus}
The study was conducted in a one-way mirror laboratory. The study area measured approximately 20 square meters and was equipped with cameras and a microphone for observation and recording. Within this space, we designated a zone for Task A that contained a standing desk (1.10 m high with a 50 cm x 50 cm surface) along with the BSK setup.
To encourage movement and varied postures during Task B, we placed three labels containing script components in different locations and at different heights throughout the room: one close to the ceiling, one below a group of chairs, and one under the table.
In both tasks, the participants used a Microsoft HoloLens 1 device equipped with transparent holographic lenses (stereo combined resolution of 1268 × 720 pixels, 60 Hz refresh rate, and a 30$^\circ$ horizontal by 17.5$^\circ$ vertical field of view). The AVAR interface~\cite{avar} was used to display and interact with the data visualizations. Text input was performed using a Bluetooth Apple Magic Keyboard.
All Task B sessions were video recorded using Noldus MediaRecorder software~\cite{noldus_mediarecorder} to support future behavioral analysis. 

\subsection{Measures}

Participants first completed a pre-test questionnaire that collected demographic information and background experience with AR. To monitor the potential adverse effects associated with AR use, we administered 14 items of the Simulator Sickness Questionnaire~\cite{kennedy1993simulator,prieto2022traduccion} after each main task (Task A1, A2, and B). The responses were recorded on a scale from 0 (no perception) to 4 (severe perception) for all listed symptoms. Performance was measured by recording the time taken to complete each transcription and the number of errors. Minor character differences that did not affect the code syntax (such as extra spaces or line breaks) were excluded from the error counts. 

To assess the subjective perception of comfort~\cite{vink2008comfort}, we used Corlett and Bishop's Postural Discomfort Questionnaire ~\cite{corlett1976technique}. This instrument presents a diagram of the human body (front and back) where participants indicate the location and severity of any discomfort on a scale from 0 (no pain) to 4 (severe pain). This instrument was also administered after each task (A1, A2, and B).

Finally, after Task B, the participants completed an additional questionnaire that evaluated the comfort of the overall BSK system and their specific features~\cite{vink2008comfort}. The questionnaire was based on previous studies exploring the usability of the product~\cite{ipaki2021study,moosburner2019real} and was adapted to the BSK. Finally, to assess general usability and acceptance, participants completed the System Usability Scale (SUS)~\cite{brooke1996sus}, which has been widely used to assess subjective perceptions of system usability.

\subsection{Procedure}

After participants provided their informed consent and completed the pre-test questionnaire, the session began with a training stage to familiarize them with the AR system. The training included an overview of the required interactive gestures and a walkthrough of the Microsoft HoloLens tutorial. The participants were then shown the keyboard commands to control the AVAR interface and gestures to manipulate the AR visualizations, including selecting, dragging, and rotating objects. Then they were asked to open an example visualization script, edit it, execute it, and manipulate the resulting visualization. Participants were encouraged to practice freely until they felt comfortable using the system.

Upon completion of the training, participants completed the Simulator Sickness Questionnaire to confirm that there were no moderate or severe symptoms before proceeding. They were also instructed to report any symptoms at any time so that the study could be stopped or paused if necessary.

For Task A, the researcher introduced a hypothetical scenario:
\emph{``Imagine that you need to supervise the operations of a bottling industry. Your objective is to review the status of the production process by writing programming code that allows you to visualize the status of machines and their productivity levels."}
The participants were then asked to transcribe a script from a printed sheet into the AVAR interface using the Bluetooth keyboard on the standing desk (A1), followed by the same activity using the BSK (A2), or in reverse order.
After each transcription task, participants removed the HoloLens headset to complete the Postural Discomfort Questionnaire to report any musculoskeletal discomfort experienced during the task. The Simulator Sickness Questionnaire was administered again after both transcription tasks were completed.

For Task B, participants were introduced to another hypothetical scenario that required them to create visualizations of factory machines and the performance of the production line. 
The researcher then demonstrated how to open a visualization script in the system and gave the participants the following instructions: (a) locate the missing piece of code in the script; (b) search the laboratory to find the label containing code segment \texttt{B1} and complete the script; (c) execute the script; (d) position the resulting graph in the designated area; and (e) repeat the process for code segments \texttt{B2} and \texttt{B3}. The researcher guided the participants through each step as needed.

After completion of Task B, the participants removed the HoloLens and completed the Postural Discomfort Questionnaire, the Simulator Sickness Questionnaire, the BSK comfort questionnaire, and the System Usability Scale to evaluate their overall experience.

\subsection{Data Analysis}

For task A, we measured both the task completion time and the number of transcription errors in each condition. In calculating errors, we excluded \del{minor} differences that \chg{did not affect the code}{are} synta\chg{x}{ctically and semantically irrelevant in Smalltalk}, such as extra spaces or line breaks. 
\ins{In particular, Smalltalk ignores whitespace as a token delimiter, and line breaks do not carry syntactic meaning; thus, these variations do not affect parsing, compilation, or execution of the code, and would not constitute a programming error in practice. 
We therefore counted as errors only deviations that could change tokens, message sends, literals, or structural elements (\eg missing or extra punctuation, incorrect identifiers, altered selectors, or mismatched brackets/parentheses), which would potentially lead to compilation/runtime failures or unintended behavior. 
This choice also improves measurement reliability by avoiding penalties for formatting differences introduced by display constraints in AR (\eg line wrapping) or individual coding styles, and by focusing the error metric on correctness rather than layout. 
}
We perform descriptive statistics on all measures and assess the normality of the data distributions using the Shapiro–Wilk test and the equality of variances using the Bartlett test before comparative analyses. Task completion times and error counts between A1 (transcription using standing desk) and A2 (transcription using the BSK) were compared using t tests when the assumptions of normality and equal variance were met.

For Task B, we conducted a behavioral analysis of the video recordings using The Observer XT 17 software~\cite{noldus_observerxt}. We coded state events, representing active behaviors sustained over a period of time, including walking, standing, standing with a flexed torso, standing with an extended torso, kneeling, sitting on the floor, lying on the floor, three neck postures (neutral, flexed, extended), and three types of hand activity (free, typing, interacting with AR). The total time spent in each posture or activity was measured. Additionally, we coded point events that do not have duration and recorded their frequency. These events included HoloLens adjustments, BSK adjustments, breaks, and expressions of discomfort.

All quantitative analyses were performed with Microsoft Excel~\footnote{\url{https://www.microsoft.com/es-cl/microsoft-365/excel}} and R software\footnote{\url{https://www.r-project.org/}}.

\section{Results}
\label{sec:results}
In this section, we present the findings of the study, including sample characteristics, quantitative performance metrics, behavioral observations, and subjective evaluations of comfort and usability. We begin by describing the demographics and background experience of the participants. We then report comparative analyses of typing performance when using a standing desk and the BSK. Next, we detail behavioral observations of movement and posture during the simulated field task, as well as the frequency of point events such as equipment adjustments and breaks. In particular, the Simulator System Questionnaire confirmed that the participants experienced no significant difficulties while using the BSK, with interruptions being exceptionally rare, only two participants paused briefly for about one minute. We also provide an overview of the postural discomfort reported in all tasks. Finally, we summarize the feedback of the participants on the perceived comfort and usability of the BSK system.


\subsection{Sample Characteristics}

Twenty adults participated in the study (mean age = 23.55, SD = 6.68). Most of the sample were students (n = 18) and had between 1 and 3 years of programming experience. Thirteen participants reported having at least basic experience with AR. Table~\ref{tab:sample-description} summarizes the characteristics of the sample.

\begin{table}[h]
\centering
\begin{tabular}{ll}
\hline
Sample description & Result \\
\hline
Age & 23.55 (6.68) \\
Gender: Male & 16 \\
Gender: Female & 4 \\
Occupation: Student & 18 \\
Occupation: Professional & 2 \\
Experience programming: Less than a year & 5 \\
Experience programming: Less than 5 years & 14 \\
Experience programming: 5 years or more & 1 \\
AR Experience: None & 7 \\
AR Experience: Basic & 11 \\
AR Experience: Medium & 1 \\
AR Experience: Advanced & 1 \\
\hline
\end{tabular}
\caption{Sample description of participants}
\label{tab:sample-description}
\end{table}

\subsection{Typing Performance (Task A)}

Participants took an average of 732.2 seconds (SD = 242) to transcribe the script using the standing desk and 848.8 seconds (SD = 279) when using the BSK. On average, the participants made 4 typing errors under standing desk conditions (SD = 3.06, range 0–10) and 6.4 errors with BSK (SD = 2.93, range 1–13).

The Shapiro–Wilk test confirmed the normality of the data distribution and Bartlett's test indicated homogeneity of the variances. The t-test analysis did not reveal significant differences in task completion time between conditions ($t = -1.99$, $df = 19$, $p = .062$). However, the number of errors was significantly higher when participants used the BSK ($t = -2.56$, $df = 19$, $p = .019$). \del{These results indicate that, while overall typing speed remained similar, accuracy was reduced when using the body-supported keyboard.

While we observed a significant increase in errors when using the BSK, task completion time was comparable to the standing desk condition. This suggests that body-supported keyboards may preserve input efficiency better than virtual or mid-air keyboards, though at the cost of reduced accuracy.

Table~\ref{tab:typing-comparison} compares the results of our study with previous works that evaluated various text entry methods in immersive environments. The observed WPM in our study (11.8--13.7) was slower than the typical typing speeds reported for plain text entry on standard keyboards (30--60 WPM), but comparable to or exceeding the speeds measured for alternative input techniques in VR and AR, such as virtual keyboards (approximately 8--16 WPM), touchscreen keyboards (around 10 WPM) and chorded keyboards (around 6 WPM)~\cite{mcgill2015dose,grubert2018text,knierim2018physical}. Compared to selection-based text entry methods such as Dasher and eye-gaze typing, which typically range from about 5 to 12 WPM~\cite{vertanen2009parakeet,majaranta2002twenty}, our body-supported keyboard also demonstrated higher or similar entry speeds. In our analysis, we calculated the WPM by dividing the total number of transcribed characters by five (to approximate the word count) and then dividing by the time in minutes required to complete the transcription. In particular, our error rates remained in the lower range relative to most of these techniques (2--4\% vs. 5--10\%), suggesting that body-supported keyboards may help preserve familiarity and accuracy while enabling mobility, although they still incur a measurable performance penalty compared to conventional desktop setups.}
\ins{Observed typing speed was 11.8–13.7 WPM. Error rates ranged from 2–4\%. WPM was computed as characters/5 divided by transcription time in minutes. We contextualize these values relative to prior AR/VR text-entry work in Section~\ref{sec:discussion}}

\subsection{Behavioral Observations (Task B)}

Participants took an average of 607.6 seconds (SD = 192.2) to complete Task B. Video analysis revealed that although most of the participants spent the majority of the time standing with their neck in a neutral position and their hands moving freely, they were able to move around the environment and change postures without difficulty. \del{This flexibility highlights the ergonomic feasibility of using the BSK for mobile authoring tasks in AR environments.}

Table~\ref{tab:behaviors-postures} details the mean time spent in different postures and activities during the task. In particular, participants spent over 400 seconds standing still on average, often alternating between interacting with AR content and typing on the keyboard. The hands were free or engaged in gestures for substantial portions of the task, reflecting the mixed demands of typing and spatial manipulation. In addition, a considerable amount of time was spent in postures closer to the ground, such as kneeling, sitting, or lying down, due to the placement of script labels at varying heights in the workspace, which encouraged dynamic movement and exploration.

\begin{table}[h!]
\centering
\begin{tabular}{lrrr}
\toprule
\textbf{Behaviors and postures} & \textbf{n} & \makecell{\textbf{Time(seconds)}} & \makecell{\textbf{Standard}\\\textbf{Deviation}} \\
\midrule
Neck - Neutral posture & 20 & 444.05 & 150.91 \\
Standing - Still & 20 & 401.95 & 132.33 \\
Hands - Moving freely & 20 & 279.90 & 90.41 \\
Hands - AR gestures interaction & 20 & 202.60 & 141.13 \\
Hands - Typing & 20 & 142.95 & 45.97 \\
Walking & 20 & 82.45 & 38.75 \\
Neck - Extended & 20 & 58.45 & 29.51 \\
Neck - Flexed & 19 & 129.32 & 84.11 \\
Kneeling & 14 & 51.79 & 43.77 \\
Sitting on the ground & 11 & 72.55 & 45.65 \\
Lying on the ground & 11 & 71.91 & 62.56 \\
Crouching & 9 & 34.78 & 30.81 \\
Standing - Flexed torso & 8 & 8.63 & 7.50 \\
Standing - Extended torso & 3 & 43.00 & 7.00 \\
\hline
\end{tabular}
\caption{Time spent in different behaviors and postures}
\label{tab:behaviors-postures}
\end{table}

\begin{figure*}[t]
    \centering
    \includegraphics[width=\textwidth]{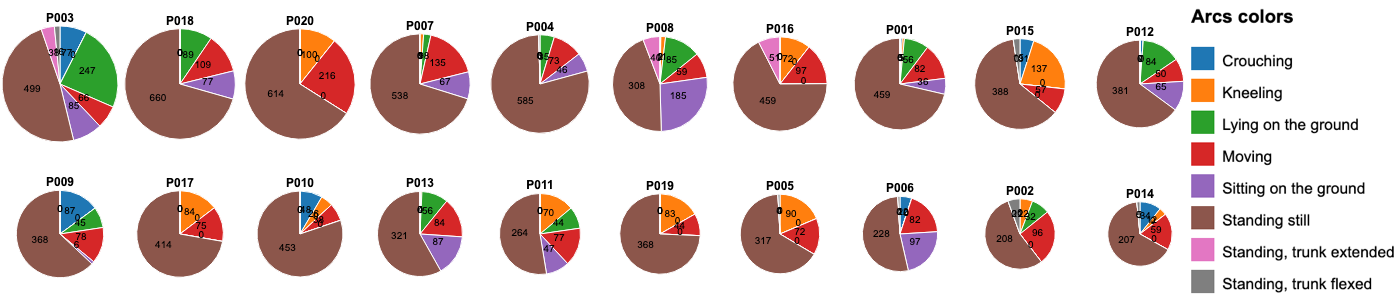}
    \caption{Body posture along the session.}
    \label{fig:body}
\end{figure*}

\begin{figure*}[t]
    \centering
    \includegraphics[width=\textwidth]{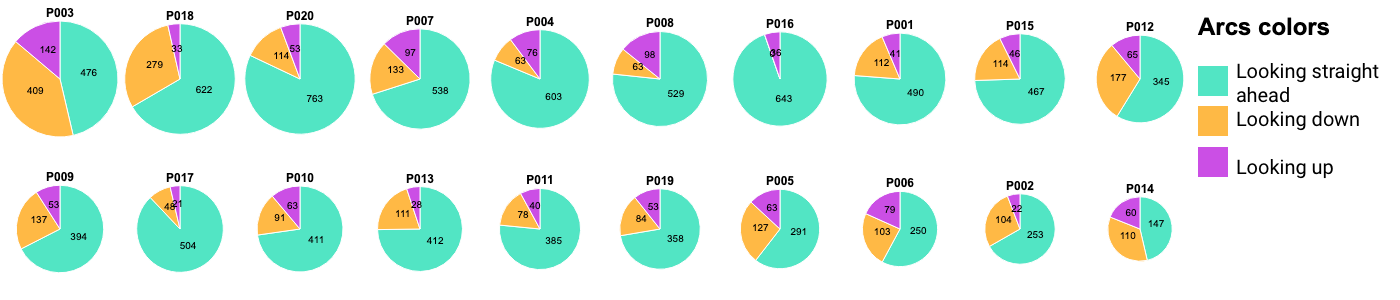}
    \caption{Head orientation along the session.}
    \label{fig:head}
\end{figure*}

\begin{figure*}[t]
    \centering
    \includegraphics[width=\textwidth]{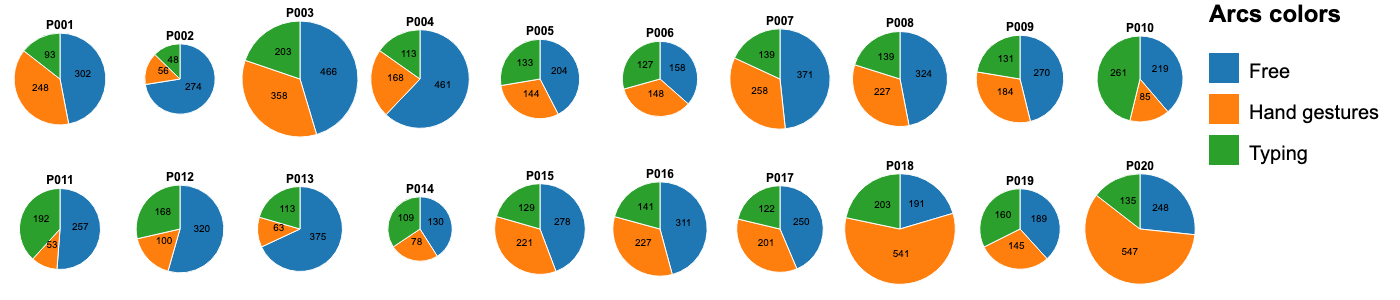}
    \caption{Usage of hands along the session.}
    \label{fig:hands}
\end{figure*}

Figures~\ref{fig:body}, \ref{fig:head}, and \ref{fig:hands} further illustrate the activities of the body, head, and hand between the participants. These visualizations show clear interindividual differences in the way participants approached the task. In particular, participants P3 and P8 were the most active, participating in a wider variety of postures compared to others, who predominantly stood still or maintained a neutral neck posture throughout the session. For example, P3 had the longest session duration and spent considerable time lying down, crouching, or sitting on the floor, likely due to interacting with low-placed script labels and repositioning AR objects. In contrast, P16 largely avoided flexing the neck and showed more static posture patterns.

Regarding the point events, Table~\ref{tab:point-events} shows the frequency of adjustments and breaks observed during task B. HoloLens adjustments were the most \chg{common, indicating the need for occasional recalibration or repositioning of the headset}{frequently observed point events}. BSK adjustments were less frequent, and only a few participants required the assistance of the researcher to adjust the system during the task. Short breaks were observed in four cases, suggesting that although the BSK supported mobility, extended use could still lead to the need for brief pauses to recover or reposition.

\begin{table}[h!]
\centering
\begin{tabular}{lccc}
\toprule
\textbf{Point Events} & \textbf{n} & \makecell{\textbf{Frequency}\\\textbf{Range}} & \textbf{Mean} \\
\midrule
Adjustment - HoloLens & 17 & 1–8 times & 3.47 \\
Adjustment - BSK & 11 & 1–3 times & 1.64 \\
Researcher support for BSK adjustment & 3 & 1 time & 1 \\
Break & 4 & 1 time & 1 \\
\hline
\end{tabular}
\caption{Point events during the study}
\label{tab:point-events}
\end{table}

\begin{figure*}[t!]
    \centering
    \includegraphics[width=\textwidth]{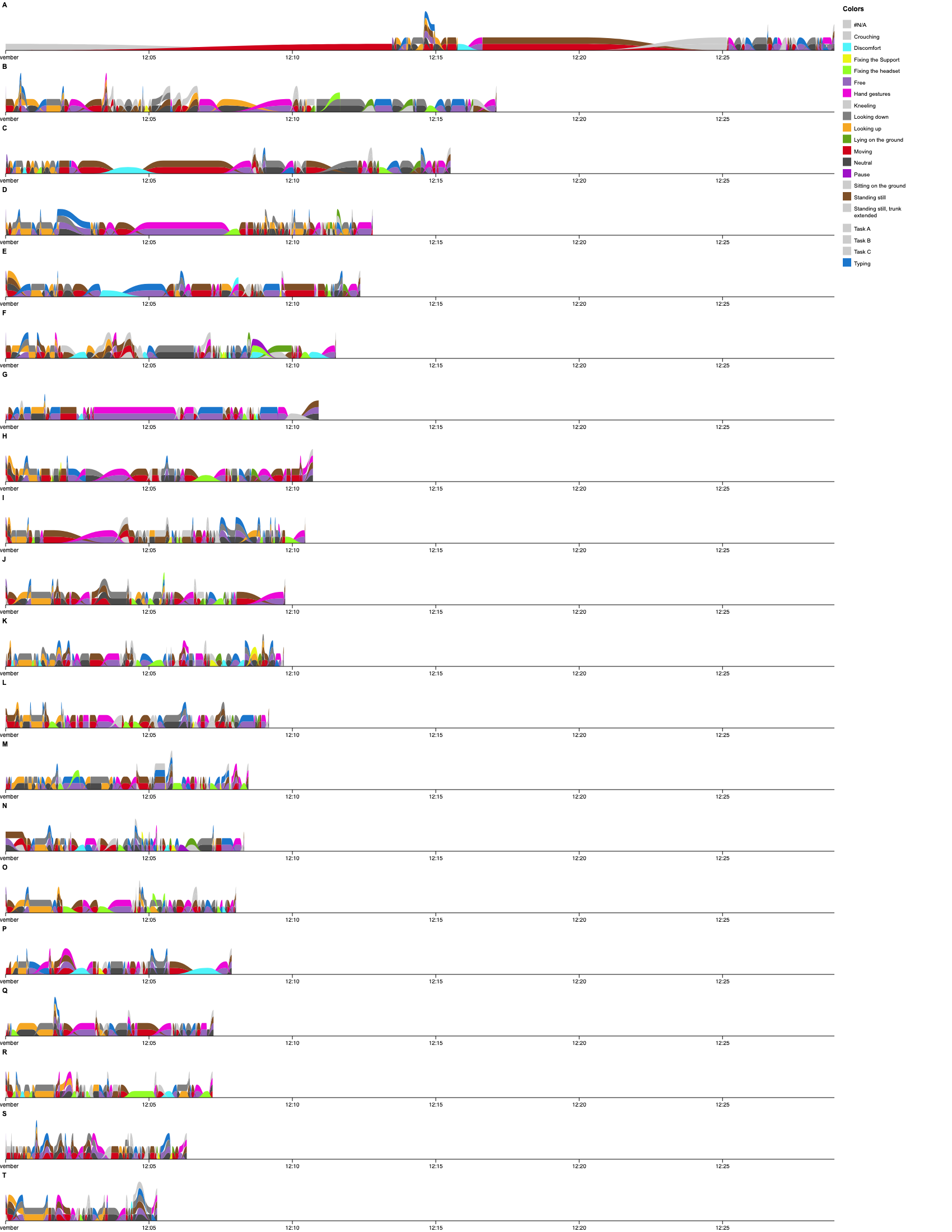}
    \caption{User behavior along the sessions.}
    \label{fig:sessions}
\end{figure*}

Figures~\ref{fig:body}, \ref{fig:head}, \ref{fig:hands}, and \ref{fig:sessions} provide visual overviews of the behavior and patterns of interaction of participants during study sessions. These figures illustrate how individuals combined movement, posture adjustments, and hand interactions to complete tasks, highlighting the diverse ways that the BSK was integrated into different working styles.

\del{Overall, these results demonstrate that while the BSK allowed participants to perform in-situ programming tasks in a variety of postures, frequent adjustments and occasional pauses were still necessary, highlighting important considerations for future improvements in comfort and system stability.}
\ins{We interpret these behavioral patterns and their ergonomic implications in Section~\ref{sec:discussion}.}

\subsection{Postural Discomfort}

Postural discomfort was assessed after each task using the Corlett and Bishop questionnaire. Table~\ref{tab:postural-pain} summarizes the number of participants who reported discomfort in each region of the body, together with the severity ratings for the three conditions.

Neck and lower back discomfort were the most frequently reported problems in all tasks. These areas often received mild to moderate ratings, reflecting the sustained periods of standing posture and head stabilization required to operate the AR headset. Some participants also noted mild discomfort in the wrists and shoulders, particularly when using the BSK to type\del{, suggesting that although the device allowed freedom of movement, extended use could still contribute to localized strain in the upper body}.

In addition, a smaller number of participants reported discomfort in the knees and ankles, likely related to kneeling or crouching to access low-placed script labels during Task B. While severe discomfort was generally uncommon, some instances of significant pain were observed, \chg{emphasizing the need for further ergonomic refinement of the body support system and task environment}{especially in the neck for Task A1 and in the lumbar back for Task A2, each reported by two participants (as shown in Table~\ref{tab:postural-pain})}.

\begin{table}[h]
\centering
\addtolength{\tabcolsep}{-0.2em}
\begin{tabular}{llccc}
\toprule
\textbf{Region} & \textbf{Severity} & \makecell{\textbf{Task A1}\\\textbf{\ins{w/} Standing Desk}\\\emph{(n participants)}} & \makecell{\textbf{Task A2}\\\textbf{\ins{w/} BSK}\\\emph{(n participants)}} & \makecell{\textbf{Task B}\\\textbf{\ins{w/} BSK}\\\emph{(n participants)}} \\
\midrule
Right shoulder & Minimum & 2 & 3 & 0 \\
Left shoulder & Minimum & 1 & 2 & 0 \\
Right elbow & Minimum & 0 & 1 & 0 \\
Left elbow & Minimum & 0 & 1 & 0 \\
Right forearm & Minimum & 0 & 1 & 0 \\
Right wrist & Minimum & 0 & 2 & 0 \\
Left wrist & Minimum & 0 & 3 & 1 \\
& Medium & 0 & 1 & 0 \\
Right hand & Minimum & 0 & 2 & 0 \\
Left hand & Minimum & 0 & 1 & 0 \\
Right knee & Minimum & 1 & 2 & 0 \\
Left knee & Minimum & 1 & 2 & 0 \\
Right ankle & Minimum & 1 & 0 & 0 \\
Neck & Minimum & 6 & 3 & 5 \\
& Moderate & 3 & 6 & 2 \\
& Significant & 2 & 0 & 0 \\
Dorsal Back & Minimum & 4 & 5 & 2 \\
& Moderate & 1 & 0 & 0 \\
& Significant & 0 & 1 & 0 \\
Lumbar Back & Minimum & 10 & 8 & 10 \\
& Moderate & 3 & 5 & 2 \\
& Significant & 0 & 2 & 0 \\
Right foot heel & Moderate & 1 & 3 & 1 \\
& Significant & 0 & 0 & 1 \\
Left foot heel & Minimum & 2 & 0 & 0 \\
& Moderate & 1 & 2 & 1 \\
\hline
\end{tabular}
\caption{Participants' reports of postural pain in Task A and Task B}
\label{tab:postural-pain}
\end{table}

\del{Overall, these results reveal that while BSK enabled mobile AR programming tasks, it did not completely eliminate musculoskeletal demands, and further ergonomic adjustments may help reduce discomfort in the most affected areas.}

\subsection{BSK Perceived Comfort and Usability}

After completion of Task B, participants evaluated the comfort and usability of the BSK system. In general, the comfort evaluations were neutral and positive in most attributes (Table~\ref{tab:bsk_usability}). Participants rated the system as fitting well to the torso and remaining stable during changes in movement and posture.  

Regarding overall usability, the BSK received a mean System Usability Scale (SUS) score of 74.5 (SD = 13.42, range 47.5–97.5)\del{, which surpasses the widely cited threshold of 68 points indicative of acceptable usability~\cite{measuringu_sus}}. The detailed responses are summarized in Table~\ref{tab:sus-responses}.\del{ These findings indicate that despite the novelty of the wearable configuration, participants generally considered the system usable and satisfactory for supporting mobile AR authoring activities}.

\begin{table*}[ht]
\centering
\footnotesize
\setlength\tabcolsep{0pt}
\begin{tabular}{>{\raggedright\arraybackslash}p{7.5cm}ccccc c}
\toprule
\textbf{Questions} & \multicolumn{5}{c}{\textbf{Frequency}} & \textbf{Mean (SD)} \\
 & Strongly Disagree & Disagree & Neutral & Agree & Strongly Agree & \\
\midrule
I feel comfortable using the BSK\\\textit{Me siento cómodo utilizando este soporte corporal para teclado} & 0 & 3 & 8 & 7 & 2 & 3.4 (0.88) \\
The BSK fits my torso well\\\textit{El soporte corporal se ajusta bien a mi torso} & 0 & 3 & 0 & 12 & 5 & 3.95 (0.94) \\
The straps fit well on my shoulders and do not cause any discomfort\\\textit{Los tirantes se ajustan bien en mis hombros y no me generan disconfort} & 0 & 1 & 0 & 14 & 5 & 4.15 (0.67) \\
Using the BSK is comfortable for my arms and hands\\\textit{Utilizar el soporte corporal para teclado es cómodo para mis brazos y manos} & 1 & 4 & 5 & 7 & 3 & 3.35 (1.14) \\
The BSK stays in the correct position while I move\\\textit{El soporte se mantiene en la posición correcta mientras me desplazo} & 0 & 2 & 1 & 7 & 10 & 4.25 (0.97) \\
The BSK stays in the correct position when I change posture or move my arms\\\textit{El soporte se mantiene en la posición correcta cuando cambio de postura o muevo los brazos} & 0 & 4 & 0 & 10 & 6 & 3.9 (1.07) \\
The BSK is too bulky or heavy\\\textit{El soporte corporal es muy voluminoso o pesado} & 8 & 6 & 3 & 3 & 0 & 2.05 (1.1) \\
The body fit system felt warm\\\textit{El sistema de ajuste al cuerpo me pareció caluroso} & 10 & 9 & 0 & 1 & 0 & 1.6 (0.75) \\
Hand interactions with the AR interface were difficult to execute due to the use of the BSK\\\textit{Las interacciones con las manos en la interfaz de realidad aumentada fueron difíciles de ejecutar debido al uso del soporte corporal} & 9 & 7 & 4 & 0 & 0 & 1.75 (0.79) \\
It is uncomfortable to type using the BSK\\\textit{Es incómodo digitar utilizando el soporte corporal} & 5 & 6 & 3 & 5 & 1 & 2.55 (1.28) \\
\bottomrule
\end{tabular}
\caption{Participant responses to comfort questions about the BSK (Body Support Keyboard) system}
\label{tab:bsk_usability}
\end{table*}

\begin{table*}[ht]
\tiny
\centering
\setlength\tabcolsep{0pt}
\begin{tabular}{>{\raggedright\arraybackslash}p{7.5cm}ccccc c}
\toprule
\textbf{System Usability Scale Questions} & \textbf{Strongly Disagree} & \textbf{Disagree} & \textbf{Neutral} & \textbf{Agree} & \textbf{Strongly Agree} & \textbf{\makecell{Participants \\ with responses \\ neutral to positive}} \\
\midrule
I think that I would like to use the BSK frequently \\
\textit{Creo que usaría este sistema de soporte corporal frecuentemente} & 3 & 3 & 4 & 9 & 1 & 14 \\

I found the system unnecessarily complex \\
\textit{Encuentro este sistema innecesariamente complejo} & 4 & 10 & 5 & 1 & 0 & 19 \\

I thought the BSK was easy to use \\
\textit{Creo que el sistema de soporte corporal fue fácil de usar} & 0 & 1 & 2 & 13 & 4 & 19 \\

I think I would need the support of a technical person to be able to use the BSK \\
\textit{Creo que necesitaría ayuda de una persona con conocimientos técnicos para usar este sistema de soporte corporal} & 8 & 11 & 1 & 0 & 0 & 20 \\

I found the various functions in this system were well integrated \\
\textit{Las funciones de este sistema están bien organizadas} & 0 & 2 & 2 & 13 & 3 & 18 \\

I thought there was too much inconsistency in this system \\
\textit{Creo que este soporte es muy inconsistente} & 5 & 9 & 5 & 1 & 0 & 19 \\

I would imagine that most people would learn to use the BSK very quickly \\
\textit{Imagino que la mayoría de la gente aprendería a usar este soporte corporal en forma muy rápida} & 0 & 0 & 2 & 11 & 7 & 20 \\

I found the system very awkward to use \\
\textit{Encuentro que este sistema es muy difícil de usar} & 9 & 9 & 2 & 0 & 0 & 20 \\

I felt very confident using the BSK \\
\textit{Me siento confiado al usar este sistema de soporte corporal} & 0 & 3 & 4 & 10 & 3 & 17 \\

I needed to learn a lot of things before I could get going with the BSK \\
\textit{Necesité aprender muchas cosas antes de ser capaz de usar este soporte corporal} & 11 & 8 & 1 & 0 & 0 & 20 \\
\bottomrule
\end{tabular}
\caption{Participant responses to the System Usability Scale (SUS) questions about the BSK system}
\label{tab:sus-responses}
\end{table*}

\section{Discussion}
\label{sec:discussion}
In this section, we reflect on the findings of our study in relation to previous work and broader design considerations. We first discuss the learning curve and performance of the BSK system compared to alternative text entry methods in immersive environments. Next, we consider how behavior, mobility, and posture shaped the user experience, along with postural discomfort and ergonomic challenges. We then examine usability perceptions and factors that influence technology acceptance. Finally, we highlight opportunities for multimodal input and summarize the contributions, limitations, and directions for future research.

\subsection{Learning Curve and Training Effect.}
Our results suggest that the BSK has the potential to offer a viable solution for text input with a traditional keyboard in AR environments. In Task A, script transcription using the BSK did not significantly increase task time compared to a standing desk. However, while participants made relatively few errors under both conditions (maximum 13), error rates were significantly higher with the BSK. This outcome likely reflects the novelty of the setup and the need for initial familiarization. Observational data showed that some participants adapted quickly to the BSK, whereas others continued to exhibit hesitations and more frequent downward glances throughout the session. Previous studies have shown that familiarity and repeated practice in immersive text entry tasks can improve speed, accuracy, and user confidence~\cite{mcgill2015dose,grubert2018text}. Future research could investigate whether structured training sessions, such as staged practice tasks or extended exposure over multiple days, reduce error rates and improve perceived usability over time.
Beyond these initial findings, previous work suggests that familiarity and practice should yield measurable improvements in both speed and accuracy for body-supported typing. In a mid-air QWERTY study, Dudley et al. observed higher entry rates by the final test block (\eg 23.4–27.6 WPM) and peak rates up to 40–45 WPM for experienced users, indicating clear gains with practice and prior familiarity~\cite{dudley2023probabilistic}. A five-session longitudinal study of a new gesture keyboard (H4VR) also reported session-over-session improvement (despite lower absolute speeds), underscoring the role of training when input mapping is unfamiliar~\cite{fallah2023h4vr}. For standard physical keyboards in VR, Grubert \etal found little learning effect because users can transfer existing typing skill~\cite{grubert2018text}; by analogy, we expect BSK improvements to come less from the acquisition of raw motor skills and more from posture optimization (fewer downward glances, smoother attention shifts) as users adapt to body support. These expectations align with established text-entry learning models (\eg power-law of-practice), which predict early gains that decrease as performance stabilizes~\cite{mackenzie1999design,mackenzie2002text}.


subsection{Comparison to Alternative Text Entry Methods.}
\ins{When comparing the BSK to existing text entry techniques for AR and VR, it is important to note that its novelty does not lie in introducing a new text entry method, but in the ergonomic, body-supported integration of a familiar physical keyboard to support mobile, in-situ authoring in augmented reality. This design choice fundamentally differentiates the BSK from mid-air, surface-projected, or selection-based approaches, and frames performance trade-offs in terms of mobility, familiarity, and ergonomic stability rather than raw typing speed alone.} 

\chg{T}{In this context, t}he observed typing speed with the BSK, averaging approximately 12--14 words per minute (WPM), was slower than conventional desktop typing speeds (typically 40--60 WPM), but comparable to or exceeding the performance reported for other immersive text entry techniques. For example, virtual keyboards in mid-air and selection-based input methods often yield WPM rates between 8 and 16, with error rates ranging from 5\% to 10\%~\cite{knierim2018physical,vertanen2009parakeet}. 
\ins{Table~\ref{tab:typing-comparison} compares the results of our study with previous works that evaluated various text entry methods in immersive environments and further illustrates this trade-off: our observed WPM (11.8--13.7) was below typical plain-text entry on standard keyboards, yet comparable to or higher than virtual keyboards, touchscreen keyboards, chorded keyboards, and other techniques evaluated in AR and VR~\cite{mcgill2015dose,grubert2018text,knierim2018physical}. Compared to selection-based text entry methods such as Dasher and eye-gaze typing, which typically range from about 5 to 12 WPM~\cite{vertanen2009parakeet,majaranta2002twenty}, the BSK also demonstrated higher or similar entry speeds.}

\ins{These results indicate that, while overall typing speed remained similar across our two conditions, accuracy was reduced when using the body-supported keyboard. While we observed a significant increase in errors with the BSK, task completion time was comparable to the standing desk condition, suggesting that body-supported keyboards may preserve input efficiency better than virtual or mid-air keyboards, though at the cost of reduced accuracy. In our analysis, we calculated WPM by dividing the total number of transcribed characters by five (to approximate the word count) and then dividing by the time in minutes required to complete the transcription.}
In contrast, our study found lower error rates (2--4\%), suggesting that body-supported keyboards can better preserve the familiarity and precision of traditional typing even in a portable configuration. This performance trade-off reinforces that while the BSK does not fully replicate the efficiency of a physical desk, it provides a promising middle ground by balancing mobility and accuracy in mobile AR scenarios. Future work could further benchmark the BSK against emerging multimodal or adaptive text input systems to assess how hybrid approaches might optimize speed, error tolerance, and user satisfaction.

\subsection{Behavior, Mobility, and Posture.}
Task B showed positive results in terms of use behavior and the ability to change posture. Although standing still was the most frequent activity overall, there was substantial variability between the participants. Some, such as P004, P010, and P014, predominantly remained upright, while others, including P003 and P018, frequently transitioned between moving, crouching, sitting, and even lying on the ground. This diversity of strategies suggests that the participants adapted their postures to maintain comfort and perform the task without major limitations. In particular, participants with longer sessions (\eg P003, P018) tended to distribute their time more evenly between multiple postures, head positions, and hand activities, indicating that prolonged use naturally increases the need for variation to mitigate fatigue. In contrast, shorter sessions (\eg P002, P014) were characterized by a higher proportion of standing still, maintaining a neutral gaze, and limited typing.
\ins{All in all, this flexibility highlights the ergonomic feasibility of using the BSK for mobile authoring tasks in AR environments.}

Importantly, standing postures with trunk flexion or extension were rare, suggesting that the BSK supported ergonomic alignment during upright work. Head orientation data showed that most of the participants spent much of the session looking straight ahead, further indicating that the display and keyboard setup promoted neutral neck posture. However, some individuals (\eg P003, P018, P001) spent extended periods looking down, probably due to frequent glances between the AR interface and the keyboard, especially during continuous typing. Data on hand usage reinforced this interpretation, showing that participants with a longer typing time often combined it with more downward head positions, while those who relied more on gestures (\eg P002, P008) maintained a more upright gaze. This pattern highlights the potential benefits of optimizing display placement or incorporating additional multimodal input to reduce neck flexion and repetitive visual shifts, especially considering that in the study, participants were reading and writing code, unlike real authoring tasks where code would be created without constant visual reference.

In addition, many participants combined periods of stillness with short bursts of movement, repositioning, and alternating hand activities. For example, P007 and P012 regularly switched between standing, walking, interacting with AR gestures, and assuming lower postures without difficulty. It is also relevant to note that participants adjusted the HoloLens device more frequently compared to the BSK itself, further suggesting that the body support did not introduce excessive interruptions or require constant reconfiguration. In general, these observations indicate that the BSK allowed participants to integrate diverse movements, head orientations, and hand input \chg{without substantial restrictions}{while performing in-situ programming tasks across a variety of postures}, a critical ergonomic consideration in field contexts where fixed postures are impractical. \ins{At the same time, the need for frequent adjustments and occasional pauses highlights important considerations for future improvements in comfort, system stability, and long-term usability.}

\subsection{Postural Discomfort and Ergonomics.}
Regarding the assessment of postural discomfort, during Task A, participants reported pain in a more diverse range of body regions when using the BSK compared to standing at the desk. This pattern probably reflects the increased muscular and postural demands of typing while wearing the body-supported system in a stationary position, which can restrict subtle weight shifts. However, this difference disappeared largely in Task B, where participants could move and change positions more freely. In all conditions, the most frequent complaints were located in the lumbar back, neck, dorsal back, and heels, areas commonly affected by prolonged standing and static load\chg{.}{, suggesting that although the device allowed freedom of movement, extended use could still contribute to localized strain in the upper body}  

Analysis of head orientation patterns showed that some participants, such as P003 and P018, spent extended periods looking down while transcribing, likely contributing to reported neck discomfort. In contrast, others such as P016 and P017 maintained a predominantly neutral gaze, coinciding with fewer neck complaints. Detailed behavioral timelines suggest that these divergent experiences were partly influenced by individual usage strategies: participants who spent prolonged uninterrupted periods typing while maintaining a downward gaze (\eg P003, P018) may have experienced greater discomfort, fatigue, or cognitive load, likely affecting their perception of comfort and usability. In contrast, those who alternated between shorter typing intervals, gestural interaction, and brief pauses (\eg P016, P017) tended to adopt more varied postures and neutral head positions, which may have contributed to more positive evaluations of comfort and acceptance of the system.  
This pattern reinforces the value of supporting dynamic workflows that integrate typing, movement, and multimodal input rather than long bouts of static text entry. 
\ins{At the same time, the presence of occasional instances of significant discomfort suggests the need for further ergonomic refinement of the body support system and the surrounding task environment, particularly for longer or more demanding authoring sessions.}
It also highlights the importance of configurable support systems that can adapt to individual work rhythms and reduce micro-disruptions over time.  

Similarly, the diversity of body postures observed in Task B, including crouching, kneeling, sitting, and lying down, appeared to help redistribute pressure on the lower back and legs. Participants who integrated these diverse positions, such as P003 and P007, appeared better able to balance static and dynamic loads, potentially mitigating discomfort compared to those who remained standing for most of the session (\eg P004 and P014).  

Overall, these results indicate that \ins{while} the BSK \chg{may be well suited for short-term}{enabled mobile programming} tasks \chg{, especially when combined with opportunities for movement and posture variation}{and supported diverse postures and movement, it did not completely eliminate musculoskeletal demands. The system appears well suited for short-term tasks, especially when combined with opportunities for movement and posture variation. At the same time, these localized discomfort patterns suggest that further ergonomic adjustments are necessary to better support the most affected areas during extended use}. However, they also highlight that even familiar devices can introduce unexpected physical demands if not carefully integrated into wearable systems. Further ergonomic improvements, such as adjustable weight distribution, breathable materials, improved load balancing, and configurable setups that allow users to adapt keyboard angle, display alignment, and posture breaks, could help mitigate discomfort during longer sessions. Future research should examine how head orientation, hand activity, and body posture interact over time to inform the design of immersive authoring toolkits that better support user well-being.

\subsection{Usability Perceptions and Acceptance.}
In general, participants were neutral to positive with respect to BSK comfort, including its adjustment to the body and stability during movement. They did not consider that the BSK was an obstacle when interacting with the AR interface or that it was excessively warm. Consequently, the mean SUS evaluation of the BSK was acceptable (mean = 74.5; SD = 13.14)\ins{, which surpasses the widely cited threshold of 68 points indicative of acceptable usability~\cite{measuringu_sus}. These findings indicate that despite the novelty of the wearable configuration, participants generally considered the system usable and satisfactory for supporting mobile AR authoring activities.}. However, it is necessary to consider that 6 participants rated overall usability with scores lower than 70, revealing that the BSK still requires design improvements to foster its adoption.

It should also be noted that the participants adjusted the BSK less frequently than the HoloLens, indicating that the body support system itself did not introduce frequent interruptions or require complex reconfiguration. From a technology acceptance perspective, the use of a familiar keyboard in AR scenarios can help facilitate adoption by reducing influential factors such as perceived ease of use, effort expectation, and performance expectation~\cite{davis1989perceived,venkatesh2003user}. 

Future studies could more systematically measure perceived usefulness, intention to use, and longer-term acceptance after repeated exposure in authentic field contexts. 
Beyond laboratory evaluation, several real-world scenarios illustrate where the BSK could be meaningfully deployed. In collaborative data analysis or design review sessions, for instance, mobile analysts could annotate or edit immersive visualizations directly in shared AR workspaces without returning to a desktop setup. In industrial maintenance or field engineering contexts, technicians often need to record sensor readings, update digital checklists, or modify scripts while standing or moving near equipment; the BSK’s body-supported design could preserve typing accuracy under such conditions. Similarly, in domains such as environmental monitoring or architecture, researchers could perform in-situ data visualization and modeling while navigating complex physical environments. These examples emphasize that portable, ergonomic text entry systems are not merely convenience tools but critical enablers of continuity between analytic reasoning, annotation, and spatial awareness in extended reality workflows. Future field trials should examine how such deployments influence collaboration, productivity, and ergonomic well-being over longer periods.

\subsection{Opportunities for Multimodal Input.}
Finally, the BSK could be complemented with other modalities such as speech recognition or hand gesture shortcuts to further reduce typing load and support individual preferences in interaction styles. The data on hand usage revealed substantial variability between participants: While some (\eg P013, P003) spent much of their session typing, others (\eg P002, P008) relied more heavily on hand gestures or kept their hands free for extended periods. This diversity highlights the need for flexible workflows that can accommodate different strategies and reduce repetitive strain. For example, integrating voice commands for frequent actions or enabling partial text input through speech could help participants who prefer to keep their hands free. Similarly, gesture-based shortcuts may allow users to complete tasks efficiently without returning to the keyboard, minimizing the need for sustained posture and frequent downward glances. Hybrid approaches that combine typing, speech, and gestures can help balance efficiency, ergonomics, and expressiveness in immersive analytics authoring. Exploring multimodal interaction frameworks could provide additional pathways to improve performance, comfort, and user satisfaction in longer sessions.

\subsection{Design Implications.}
The findings of this study point to several considerations for the design of ergonomic, expressive authoring toolkits in immersive analytics.

\begin{itemize}
    \item \textbf{Support for Dynamic Posture Variation:} Participants benefited from the ability to transition between standing, walking, and lower postures (\eg crouching or sitting), which helped mitigate discomfort during longer sessions. Systems should explicitly accommodate and encourage movement and repositioning rather than assuming static standing use.
    
    \item \textbf{Display and Input Alignment:} Extended downward glances were associated with increased neck discomfort in several participants (\eg P003, P018). Future designs should explore adjustable display placement, configurable keyboard angles, or head-up overlays that reduce the need for frequent gaze shifts between input and output.
    
    \item \textbf{Multimodal Interaction Integration:} The observed diversity in hand use strategies highlights the importance of offering flexible workflows that blend typing, speech, and gestures. Integrating voice commands for repetitive commands and gesture shortcuts for common tasks could help users reduce reliance on prolonged typing and adapt to individual preferences.
    
    \item \textbf{Ergonomic Load balance:} Even familiar devices, such as Bluetooth keyboards, can introduce unexpected musculoskeletal demands when integrated into wearable systems. Designers should prioritize adjustable weight distribution, breathable materials, and mechanisms that distribute load across larger body areas to reduce localized strain.
    
    \item \textbf{Configurable Support and Training:} Participants who kept their typing continuous often reported more frequent discomfort and adjustments. Providing configurable break reminders, customizable support configurations (\eg, tray height, angle), and structured training protocols can help users acclimate more comfortably and sustainably over time.
\end{itemize}

These recommendations can inform the development of next-generation immersive analytics toolkits that balance expressiveness, mobility, and ergonomic safety in real-world scenarios.

\section{Threats to Validity}

Our study provides an initial demonstration of the feasibility of using wearable keyboard support to program immersive analytics visualizations in AR. However, several limitations must be considered both regarding the design of the BSK prototype and the controlled evaluation methodology.
\vspace{0.3cm}

\noindent\textbf{Design and Implementation Limitations.}
The BSK was developed as an initial functional prototype with an emphasis on ergonomics and basic usability. Consequently, certain implementation choices may have influenced participants’ experiences.

\noindent\emph{Transportation and Setup:} Carrying, securing, and removing the BSK required minor adjustments and occasional assistance, which may hinder adoption in workflows that require frequent transitions between tasks, collaboration, or movement between locations.

\noindent\emph{Structure and Fit:} To promote correct posture and keyboard stability, the vest uses a snug fit that can limit the ability of users to handle additional tools or equipment. While this design improves short-term stability and comfort, it could restrict users when performing physically demanding or multi-device tasks. Future versions may benefit from adjustable harnesses and more flexible materials.

\noindent\emph{Material Properties:} The current prototype relied on standard materials that may not be breathable or lightweight enough for extended use in field conditions. Further refinements should focus on optimizing weight distribution, ventilation, and thermal comfort, especially for mobile or outdoor use.
\vspace{0.3cm}

\noindent\textbf{Controlled Evaluation Limitations.}
The evaluation setting and the characteristics of the participants impose important constraints on the generalizability of the findings.

\noindent\emph{Sample Size and Demographics:} The study included only 20 participants, mainly students and young adults with limited experience in AR and professional field work\ins{, and the gender distribution was unbalanced}. While this demographic does not represent the full range of target users, such as engineers or technicians, \chg{p}{it is important to note that age, gender, and long-term professional experience may influence typing strategies, posture preferences, fatigue perception, and ergonomic tolerance. P}rior studies in human–computer interaction have shown that student participants can provide reliable insights for early-stage usability and ergonomic evaluations when domain-specific expertise is not essential~\cite{nielsen1994usability,clemmensen2009cultural,salman2015students}. Therefore, the results should be interpreted as \chg{initial}{controlled} feasibility evidence\ins{that supports internal validity,} rather than as conclusive validation across professional populations.

\noindent\emph{Ecological Validity:} While Task B simulated mobility and interaction, the study was conducted in a structured, low-risk laboratory environment. As a result, it did not fully capture typical challenges of real-world AR scenarios, such as uneven terrain, distractions, or time pressure, which could affect both usability and ergonomics.

\noindent\emph{Short-Term Exposure:} Participants interacted with the BSK only for limited sessions. Longer-term use could reveal different adaptation patterns, cumulative fatigue, or ergonomic adjustments that were not captured within this experiment.

\noindent\emph{First-Time Use and Learning Effects:} All participants were first-time users of BSK. Higher error rates and adjustments during typing may reflect a lack of familiarity rather than intrinsic design flaws. Although the order of tasks was counterbalanced to mitigate learning effects, future longitudinal research should examine how comfort and performance evolve with continued use and training to better capture real adoption dynamics.
\vspace{0.3cm}

Taken together, these limitations highlight the need for iterative user-centered design processes and longitudinal field studies to refine the BSK system and rigorously assess its effectiveness across diverse contexts, tasks, and user populations. Overall, the findings should be understood as exploratory feasibility evidence that informs subsequent design iterations and long-term ergonomic validation.

\section{Conclusion}
In this study, we introduced an ergonomic Ergonomic Body-Supported Keyboard (BSK) system designed to enable programming immersive analytics visualizations directly within augmented reality environments. Through a controlled experiment involving stationary and mobile tasks, we evaluated the performance, physical comfort, and perceived usability of the system.

Our findings indicate that while text input with the BSK resulted in a statistically significant increase in errors compared to using a standing desk, it did not have a substantial impact on task completion time, and the overall error rate remained lower than those reported for other input devices designed for AR environments. In particular, participants were able to move freely, adopt varied postures, and interact naturally with the AR interface, demonstrating that the BSK supports mobility and flexibility in immersive settings. Postural discomfort reports were more frequent during stationary tasks but decreased when participants were able to change posture during mobile tasks, underscoring the importance of dynamic movement in ergonomic design.
Subjective feedback on the BSK was generally neutral to positive and the mean SUS score of 74.5 suggests that the system was perceived as usable. These results support the potential of integrating traditional input devices into AR workflows when combined with ergonomic supports centered on the body. However, improvements in weight distribution, and stability are needed to enhance comfort and usability during prolonged use.

In sum, this work contributes to the development of situated visualization tools by addressing a critical limitation in text input. It highlights the value of embedding physical ergonomics into the design of immersive computing systems to promote usability, reduce strain, and support the practical deployment of AR technologies in professional contexts.

Future research should pursue more iterative design cycles and involve a broader range of users, including industrial fieldworkers and individuals with accessibility needs, to create more inclusive and adaptable solutions. 
In particular, in the future, we plan to expand the BSK framework toward multimodal and physiological integration. Combining the BSK with lightweight sensing (\eg eye-tracking, EMG, or heart-rate monitoring) could provide insight into user fatigue, workload, and ergonomics during extended AR authoring sessions. Comparative studies with complementary input modalities, such as speech or gesture-based text entry, would help clarify how hybrid interaction strategies can balance efficiency, comfort, and expressiveness in different contexts. Together, these directions point toward richer, adaptive AR authoring toolkits that align physical comfort with expressive, collaborative interaction.
Such efforts can help ensure that systems like the BSK evolve into robust and versatile tools capable of supporting diverse practices in real-world environments.


\section*{Disclosure statement}
We have no interests to declare.

\section*{Data availability statement}
The data that support the findings of this study are openly available in figshare at \url{https://doi.org/10.6084/m9.figshare.29582126}, reference number 29582126.

\section*{Funding}
This work was supported by the ANID FONDECYT Iniciación under Grant 11230349; and ANID Proyectos de Exploración under Grant 13250116.

\section*{ORCID}
Leonel Merino\,\orcidlink{0000-0002-5396-487X} \url{https://orcid.org/0000-0002-5396-487X}\\
Begoña Juliá-Nehme\,\orcidlink{0000-0003-4888-9868} \url{https://orcid.org/0000-0003-4888-9868}\\
Santiago Viana\,\orcidlink{0009-0000-9343-9095} \url{https://orcid.org/0009-0000-9343-9095}


\section*{Author contributions}
CRediT: \textbf{Leonel Merino}: Conceptualization, Methodology, Formal analysis, Investigation, Supervision, Writing -- original draft, Writing -- review \& editing; \textbf{Begoña Juliá-Nehme}: Conceptualization, Methodology, Formal analysis, Investigation, Supervision, Writing -- original draft, Writing -- review \& editing; \textbf{Santiago Viana}: Investigation, Data curation, Writing -- review \& editing.

\bibliographystyle{tfp}
\bibliography{keyboard}
\section*{Notes on contributors}

\noindent\textbf{Leonel Merino} is an Assistant Professor of Engineering Design at the School of Design and School of Engineering, Pontificia Universidad Católica de Chile. His research interests include software engineering, information visualization, XR, and HCI. He holds a Ph.D. in Computer Science from the University of Bern. Contact: leonel.merino@uc.cl\\

\noindent\textbf{Bego\~{n}a Juli\'{a}-Nehme} is an Assistant Professor at the School of Design and School of Engineering, Pontificia Universidad Católica de Chile (UC). Her research interests include applied ergonomics, HCI, UX, usability, user-centered design, and inclusive design. She holds a Ph.D. in Psychology from UC and a Masters degree in Ergonomics from Universitat Politècnica de Catalunya. Contact: mbjulia@uc.cl.\\

\noindent\textbf{Santiago Viana} is a student at the School of Design, Pontificia Universidad Católica de Chile. His academic interests include UX design, human-centered innovation, and the use of emerging technologies, such as virtual and augmented reality, in design processes. He has participated in interdisciplinary projects on user interaction. Contact: sviana@uc.cl.\\

\end{document}